# ASTROMOVES: Astrophysics, Diversity, Mobility

Jarita C. Holbrook
Science, Technology & Innovation Studies,
University of Edinburgh
United Kingdom
Jc.holbrook@ed.ac.uk

## Abstract

The US astronomy/astrophysics community comes together to create a decadal report that summarizes grant funding priorities, observatory & instrumental priorities as well as community accomplishments and community goals such as increasing the number of women and the number of people from underrepresented groups. In the 2010 US National Academies Decadal Survey of Astronomy (National Research Council, 2010), it was suggested that having to move so frequently which is a career necessity may be unattractive to people wanting to start a family, especially impacting women. Whether in Europe or elsewhere, as postdocs, astrophysicists will relocate every two to three years, until they secure a permanent position or leave research altogether. Astrophysicists do perceive working abroad as important and positive for their careers (Parenti, 2002); however, it was found that the men at equal rank had not had to spend as much time abroad to further their careers (Fohlmeister & Helling, 2012). By implication, women need to work abroad longer or have more positions abroad to achieve the same rank as men. Astrophysicists living in the United Kingdom prefer to work in their country of origin, but many did not do so because of worse working conditions or difficultly finding a job for their spouse (Fohlmeister & Helling, 2014). In sum, mobility and moving is necessary for a career in astrophysics, and even more necessary for women, but astrophysicists prefer not to move as frequently as needed to maintain a research career. To gather more data on these issues and to broaden the discourse beyond male/female to include the gender diverse as well as to include other forms of diversity, I designed the ASTROMOVES project which is funded through a Marie Curie Individual Fellowship. Though slowed down by COVID-19, several interviews have been conducted and some preliminary results will be presented.

# Introduction

The ASTROMOVES project is to study the career decision making of astrophysicists and those in connected fields, the changes of positions in terms of relocations and changes in job title while remaining in the same place, and if and how these are related to intersectional identities. 'Intersectional identities' attempts to capture the many axes of difference embodied by individual astrophysicists and is in reference to 'intersectionality' defined as:

> *The interconnected nature of social categorizations such as race, class, and gender, regarded as creating overlapping and interdependent systems of discrimination or disadvantage (Oxford English Dictionary, n.d.)*

When considering intersectional identities there is space to consider nationality, mother tongue and disabilities/abilities along with race, class and gender as mentioned in the definition. Astrophysicists have long been tackling how to become more diverse in terms of including women and members of underrepresented groups (Cowley et al., 1974). Most of the social science studies of astrophysicists have focused on women; to summarize the findings of a few of these studies women and men both need to relocate multiple times for postdocs but the outcome, securing a permanent position, is different for women and men. Every ten years, there is an USA based exercise where the larger astrophysics community comes together to assess the current state of the discipline and to lay out the priorities for the coming decade; the resulting report is referred to as the Decadal Survey. In the 2010 Decadal Survey (National Research Council, 2010), it was suggested that having to move so frequently which is a career necessity may be unattractive to people wanting to start a family, especially impacting the careers of women. This suggestion is supported in part by the American Physical Society's Ivie & White (2015) study which showed that women in physics & astronomy associate having children with negatively impacting their career and slowing their rate of promotion. Also supported in part by Fohlmeister & Helling (2012) study which found that "Those who became mothers feel very restricted in mobility and point out that it is harder to combine job and family." In that same study, the second most cited reason for moving was to be near family. These findings point to conflicts between the need to relocate often for career advancement and the needs of women to have stability and external support to maintain a family and career. Men astrophysicists face the same family issues but they rate the importance of these issues to their careers as much lower than the women astrophysicists. Another issue is where astrophysicists relocate to; in the European context; solar astrophysicists perceived working abroad, that is working outside of their home country, as important and positive for their careers (Parenti, 2002).

However, Fohlmeister & Helling (2014) found that the men at equal rank had not had to spend as much time abroad to further their careers. By implication, women need to work abroad longer or have more positions abroad to achieve the same rank as men who have worked abroad. Astrophysicists living in the United Kingdom prefer to work in their country of origin, but many did not do so because of worse working conditions or difficultly finding a job for their spouse (J. Fohlmeister & Helling, 2014). Summarizing these research results for studies of women in astrophysics and their careers, moving is necessary for a career in astrophysics, and working abroad is even more necessary for women, but there is some evidence that astrophysicists prefer not to move as frequently as needed to maintain a research career. There are a few studies of intersectionality and astrophysicists such as Ko et al. (2013), which found that women of color in some cases prioritized life work balance in general, including family life, over remaining in their chosen field and revealed the positive role that activism plays, as well as importance, in women of color's lives even when doing activism is not academically rewarded and not considered as part of advancement. This and other studies considering intersectional identities in astrophysics are not focused specifically on career moves.

ASTROMOVES' focus on mobility and career decision making and its use of qualitative interviews with astrophysicists provides a deep dive towards surfacing previously unidentified factors as well as greater comprehension of known factors important for navigating astrophysics careers. One of the goals of ASTROMOVES is to leverage the research findings to make recommendations to the astrophysics community for retaining and promoting their diverse members. Though in its beginning stage, the project has preliminary results that reveal decision making factors previously not recorded, identified a new gender related outcome and has expanded previous studies by including greater gender diversity.

# Population

ASTROMOVES is a study of astrophysicists and scientists in adjacent fields such as space scientists, planetary scientists, etc. Each person had to be at least two career moves past their doctorate. Given that the project was funded with European monies, each person had to have lived, worked or studied in Europe at some point in their career. Targeted populations include people that are ethnically and nationally underrepresented in astrophysics, gender diverse people, people with physical disabilities, and people with invisible disabilities or different abilities such as dyslexia or an autoimmune disorder. Ethically, everyone who chooses to participate in ASTROMOVES volunteers and they have a say in how their data is used as well as how it appears in print.

The ASTROMOVES initial population of seven scientists was divided into four males and three females, but their self-reported gender identities include two bisexuals, one homosexual and one asexual, thus over half are embodying gender diversity. Gender diverse astrophysicists were identified using the Astronomy & Astrophysics Outlist (see [https://astro-outlist.github.io](https://astro-outlist.github.io) (Mao & Blaes, n.d.)]. Among the seven scientists are people of African descent, Indian descent, Middle Eastern descent, and European descent. Snowball sampling was used, where people were asked to recommend other people for ASTROMOVES interviews, and so on.

## Data Collection and Analysis

The primary method of data collection is interviews of between 30 minutes to 1.5 hours depending upon how many career moves the person being interviewed has made. These qualitative interviews cover the details of each career move and their decision making, as well as including places they did not apply to and offers they did not accept. Each interview was video recorded or voice recorded and transcribed. As a social political statement reminding astrophysicists of their debt to indigenous communities for allowing telescopes to exist on their sacred mountaintops, each person is given a Hawaiian pseudonym which is used across ASTROMOVES publications.

Publicly available documents are used to contextualize the interviews. To get the sequence of career moves, publicly available CVs are primarily used. When a CV is not available, the career sequence can be approximated by noting the affiliations given for their scientific articles found on the Astrophysical Data System archive (https://ui.adsabs.harvard.edu/), which is publicly available, also. The Astrophysics Rumour Mill (Anonymous, 2020) is another public source of information about job offers that is fairly complete back to 2000. These documents are used as prompts for the interviews.

Sensitive information related to intersectional identities was recorded or not depending upon the desires of the interviewee. Such information was written down and sent after the interview; interviewees had the option to write down any additional information that they wanted to keep private but that they thought was important for the analysis. In addition to personal information about their intersectional identities, the astrophysicists were hesitant to name departments with negative reputations or bad reputations, such as those that treated women postgraduate students badly by not supporting them or graduating them; this information was written as private information.

Qualitative analysis software was used to analyse the contents of the interviews. To anonymize information on specific institutions, their Shanghai designations for physics, which includes astrophysics, were used (*Shanghai Ranking's Global Ranking of Academic Subjects 2020 - Physics | Shanghai Ranking - 2020*, n.d.). Repeated themes, topics of

discussion, and experiences were compared across the set of interviews. Given the small sample size of seven interviews all results presented in the following section are preliminary.

## Results and Discussion

> *"You know, it was a different environment than I was used to. It was perhaps a little more (regional stereotype on being more formal). [USA Top 50 location] is very laid back. There it's very, it's very congenial. There's a lot of interaction between faculty and students and postdocs, and it's feels like more of a community. [USA Top 10 location] is a little more hierarchical. And it was, you know, there was the top people and then there was, you know, others. And, yeah, I didn't, I felt like I didn't, there were some people...You know, individually, I liked many of the people there, but the overall atmosphere I felt was more sort of competitive and not so friendly. And, so I, I wasn't that sad to leave. You know. And I had visited the [2nd USA Top 50 location] and I knew that …some of the people there. So, I knew I would have…like a nice community there."*

The above except from one of the interviews is a true transcription of what was said without correcting for filler words nor repetitions. When the interviewee was asked about attaching their name to this quote, the answer was only if it were grammatically corrected, thus it is presented anonymously. As for the content, many of the astrophysicists spoke about the academic environments that they preferred or didn't. Only one person stated that they enjoyed and thrived in competitive environments, most indicated that they preferred less competitive environments similar to the quote above.

> *"The end of my most recent research fellowship, then three months of unemployment and it was pretty harrowing: I'VE NEVER BEEN UNEMPLOYED BEFORE!"*

All three of the females interviewed have been unemployed at some point since earning their doctorate. Two of the three were recently unemployed connected to COVID-19. During their period of unemployment, two had to live on their own personal savings since they were not eligible for public unemployment funds; the third was eligible for and used some public unemployment assistance funds. In contrast, none of the four males interviewed had been unemployed since their doctorate. The fact that 100% of the females had experienced unemployment indicates greater job insecurity; in fact, one person felt strongly that if they had been male, then the situation which lead to unemployment would not have happened.

If female astrophysicists do have higher job insecurity, then their supervisors should consider securing additional funds to continue to employ them until they obtain a new position. Two of the male astrophysicists mentioned having positions made with them in mind, but neither was on the verge of unemployment at the time, rather the positions were made to attract them to a new institution.

> *Kamea: "I've actually got to [the] interview stage for a couple of jobs and, and they were just unceremoniously called off…months later they called me back and said, 'We'd like to resume the interview process with you.'"*

> *Maka'ala: "I'm not happy about how this has affected my, you know, my life and my ability to be with people that I care about…in terms of my productivity, it has had no effect or even a positive effect, because all the things like travel and hanging out with friends or, you know, just whatever that I would do, I don't have any more. So, I have fewer distractions."*

The females' experiences of unemployment cannot be disentangled from the COVID-19 pandemic. To disentangle this, future interviews would have to be done with more astrophysicists that were on the job market during 2020 as well as interviews with other astrophysicists that experienced unemployment prior to 2020. Checking the Astrophysics job page in the fourth quarter of 2020 (Anonymous, 2020), of the 11 jobs that mention COVID-19, 2 temporarily suspended their searches, 7 cancelled the job and 1 offer was rescinded. The quote by Kamea shows how Kamea was part of the group of astrophysicists that were effected by the astronomical job market's fluctuations due to COVID-19.

Maka'ala is in a permanent position, so COVID-19 did not challenge their job security; rather, Maka'ala speaks of the isolation brought about by the pandemic. Most of the astrophysicists that lived alone were unhappy about their isolation to the point that one, Hema, self-described as an extrovert, regularly did meet people socially even thought it was against the local government regulations.

> *Hema: "It's not so much about the number of persons, a day that I need to speak [to]. It's more about the time of day…I spend the entire day running around, because I have a job. I've work to do. I have training in the evening after work, which requires a lot of my energy - physical energy. But if I come home at 7 pm. And I don't have a plan of seeing someone that's what makes me feel really lonely…I*

> *can talk to people at work all day. But if I'm home alone in the evening with nothing to do. That's the hard part for me. That's the vulnerable moment for me, it's sort of being home alone in the evening with no one to talk to…"*

Hema and another astrophysicist, Haoa, were both given permission to work in their place of employment during COVID-19. Hema requested permission; whereas Haoa didn't have a choice because of Haoa's leadership role in an observatory related construction project.

The combination of teaching online and doing research was having a negative impact on Maka'ala:

> *"there's just a lot to keep up with and I can't, you know, if it's…if it's…if I'm working on a paper I can always say, 'No, I'm just going to stop at this point and take a break,' but I can't…I feel like I've lost my agency in terms of setting my schedule. Even if I, even if I might still be working long hours in the summer when I was doing research versus now when I'm teaching so…that's been a little bit mentally tiring."*

It should be noted that a few articles have been published on how COVID-19 is exacerbating the differences between the careers of women and men in astrophysics, such as Inno et al. (2020) and Venkatesan et al. (2020). Both articles point out the drop in publication submissions from women during the pandemic will have the long term impact of women being less competitive, because currently astrophysics competitiveness is attached to publication rates. They advocate that the pandemic impact be considered in future hiring decisions.

Contemplating unemployment and the females interviewed, financial considerations have to be considered among the factors important for career decision making in astrophysics. Postdocs and other fellowships come with a fixed salary, which means everything is equal across the sexes. However, in most cases for permanent positions, salaries are negotiable as are what are called 'start-up packages', the onus is on the new faculty member to ask for what they need to be successful. In their study of women scientists that had taken a career break, which included astronomers, Mavriplis et al. (2010) found that women were the least confident about negotiating for their salary and start-up package. This initial negotiation, which women are not confident about, may be the start of building resentment and discontent if and when they discover that others have negotiated for and gotten more both before and after they were hired. Finances were discussed during the interviews and most said that they did not make career decisions based on salary or cost of living. However, some did note when they had taken a pay cut. One of the women did provide information about negotiating for salary and start-up, which is still being analysed. None of the females had purchased homes; whereas three of the four males had purchased homes in the place where they thought they would remain, even though they later may have moved. If this same information is considered in terms of marital status, two of the males that purchased homes were married; however, these two were the only married people among the astrophysicists interviewed.

## Conclusions

ASTROMOVES, though still in its beginning phase, has produced interesting results that have to be tempered by the small number of interviews completed. Nonetheless, the most troubling findings are related to sex where the female astrophysicists had all experienced unemployment post PhD. Generally, the astrophysicists did not consider salary, cost of living or start-up package in their career decision making. COVID-19 has changed the lives of astrophysicists; many of those interviewed that were living alone spoke of their unhappiness connected to loneliness, and two were impacted due to the fluctuating job market due to COVID-19.

As ASTROMOVES continues, the interview pool will be increased to 50 scientists; an effort is being made to include those with physical disabilities and invisible disabilities. Already, there is enough gender diversity to have three gender categories for future analysis: cis-men, cis-women and other gender; which will be the first time such categories are possible in discussions of the mobility and career decision-making of astrophysicists.

## Acknowledgements

This research has made use of NASA's Astrophysics Data System Bibliographic Services. This project is funded by the European Union's Horizon 2020 research and innovation programme under the Marie Skłodowska-Curie individual fellowship program H2020-MSCA-IF-2019 grant #892944.